\documentclass[aps,preprint,amsmath,amssymb,nofootinbib]{revtex4}
\usepackage {lscape}
\usepackage{graphicx}
\usepackage{amsmath}
\usepackage{float}
\usepackage{array}
\usepackage{subfigure}

\usepackage[colorlinks,
bookmarks=true,
linkcolor=blue,
urlcolor=blue,
anchorcolor=black,
citecolor=blue
]{hyperref}

\begin{document}
	
\title{Two-dimensional excitons in monolayer transition metal dichalcogenides from simple models and variational calculations}
	
\author{J.-Z. Zhang}
 \email{phyjzzhang@jlu.edu.cn}
\author{J.-Z. Ma}%
 \affiliation{School of Physics, Jilin University, Changchun, 130012, China.
}

\date{\hfill \today}

\begin{abstract}

Exciton spectra of monolayer transition metal dichalcogenides (TMDs) in various dielectric environments are studied using an effective mass model incorporating a screened two-dimensional (2D) electron-hole interaction described by the Keldysh potential. 
Exciton states are calculated by solving a radial equation (RE) with a shooting method including Runge-Kutta integration. 
Particular attention is paid to the simple models for 2D exciton calculation. 
The 2D hydrogen model yields much lower exciton energies than the Rydberg series from the RE solution. 
 The screened hydrogen model (SHM) [Phys. Rev. Lett. {\bf 116}, 056401 (2016)] is examined by comparing its exciton spectra with the RE solutions.   
While the SHM is found to describe the nonhydrogenic exciton 
Rydberg series (i.e., the energy's dependence on main quantum number $n$) reasonably well, it fails to account for the linear decrease of the exciton energy with the orbital quantum number $m$. The exciton Bohr orbit shrinks as $\lvert m\rvert$ becomes larger resulting in 
increased strength of the electron-hole interaction and a decrease of the exciton energy. 
The exciton effective radius expression of the SHM can characterize the exciton radius's dependence on $n$, but it cannot properly describe the exciton radius's dependence on $m$, which is the cause of the SHM's poor description of the exciton energy's $m$-dependence.  
For monolayer WS$_2$ on the SiO$_2$ substrate,   
our calculated $s$ exciton Rydberg series agrees closely with that measured by optical reflection spectroscopy [Phys. Rev. Lett. {\bf 113}, 076802 (2014)],  
while the calculated $p$ excitons offer an explanation for the two broad features of a two-photon absorption spectrum [Nature {\bf 513}, 214 (2014)]. 
Our calculated exciton energies for monolayer TMDs in various dielectric environments compare favourably with experimental data. 
Variational wave functions are obtained for a number of strongly bound exciton states and further used to study the Stark effects in 2D TMDs, 
an analytical expression being deduced which yields a redshift of the ground state energy to a good accuracy. 
The numerical solution of the RE combined with the variational method provides a simple and effective approach for the study of 2D excitons in monolayer TMDs. 
	
\end{abstract}

\maketitle
\section{Introduction}

Photoexcitation creates bound electron-hole ({\it e-h}) pairs, namely, excitons, in a direct bandgap semiconductor \cite{Haug:2004}. 
A series of exciton energy levels is usually referred to as an exciton spectrum \cite{Huangst:2013,Olsen:2016}. 
Strongly bound excitons have been predicted theoretically \cite{Komsa:2012,Qiudy:2013} and also observed experimentally \cite{Yez:2014,Ugeda:2014} in monolayer (ML) transition metal dichalcogenides (TMDs), an important class of  two-dimensional (2D) semiconductors with potential for optoeletronic and valleytronic devices \cite{Wangqh:2012,Makf1:2012}. These 2D excitons have attracted intense research interest as they play a key role in strongly enhanced photoluminescence \cite{Mak:2010,Splendiani:2010} and photocurrent generation \cite{Klots:2014}, and are also involved in the excitonic absorption and recombination of valley polarization (i.e., valley-selective circular dichroism \cite{Makf1:2012,Cao:2012}). Hence, knowing the exciton spectrum is fundamental to the study of strong light-matter interactions in these 2D semiconductors \cite{Chernikova1:2014,Yez:2014,He:2014}.

Unlike bulk semiconductors, 2D semiconductors such as ML TMDs have a dielectric function $\varepsilon(\mathbf{q})$ linearly dependent on wavevector $\mathbf{q}$, $\varepsilon(\mathbf{q})=1+2\pi\alpha_{2D}q$ (Ref.\cite{Cudazzo:2011}), where $\alpha_{2D}$ is the 2D polarizability of the monolayer, leading to dielectric screening that is nonlocal in real space. The nonlocal screening has a twofold influence on the excitonic energy levels.  First, the reduced screening in two dimensions enhances  the {\it e-h} interaction making excitons in TMDs have a large binding energy, ranging from several tenths of an eV to one eV for the ground state, which is much greater than in bulk semiconductors or semiconductor quantum wells (typically on the order of 0.01 eV). Second, the screened  
{\it e-h} interaction \cite{Keldysh:1978,Cudazzo:2011} due to the dielectric function $\varepsilon(\mathbf{q})$ differs from the usual 2D Coulomb interaction, and thus 2D exciton energy levels are expected to deviate from the Rydberg series, i.e., the energy's dependence on main quantum number $n$, of a 2D hydrogen model (2DHM). This prediction has been experimentally confirmed by several experimental studies, with optical spectroscopy measurements performed on ML WS$_2$ on a SiO$_2$ substrate \cite{Chernikova1:2014,Yez:2014} and also for ML WSe$_2$ on a SiO$_2$ substrate \cite{He:2014}. Ground state exciton binding energies have been measured for a freely suspended ML of 
MoS$_2$ \cite{Klots:2014} and TMD monolayers in a variety of dielectric environments, such as MoS$_2$ on substrate SiO$_2$ \cite{Hill:2015,Rigosi1:2016} or encapsulated in hBN \cite{Robertc:2018}, as well as MoSe$_2$ \cite{Liuhj:2015}, 
WS$_2$ \cite{Chernikova1:2014,Yez:2014,Hill:2015,Rigosi1:2016,Stier:2016,Zhubr1:2015}  
and WSe$_2$ \cite{Huangja1:2016,He:2014,Wangg2:2015,Liuhj:2015,Diware:2018,Hanbicki:2015}, all on a SiO$_2$ substrate.

Exciton energies of freestanding ML TMDs have been calculated using various methods of band structure calculation and models of the electron-hole interaction.  
A rigorous treatment of excitons is the use of the Bethe-Salpeter equation (BSE) based on a first-principles GW quasiparticle band structure \cite{Feng:2012,Ramasubramaniam:2012,Shih:2013,Qiudy:2013,Olsen:2016}.   
There has also been a treatment using an effective mass model in conjunction with a first-principles calculation of the screened {\it e-h} interaction \cite{Andersen:2015,Olsen:2016}. A further simplified approach is the use of an analytical expression for the screened {\it e-h} interaction such as the Keldysh potential \cite{Keldysh:1978,Cudazzo:2011}  whilst combining it with a band-structure model such as a tight-binding \cite{Berghauser:2014}, effective near band-edge Hamiltonian \cite{Konabe:2014}, or effective mass \cite{Berkelbach:2013,Kylanpaa:2015,Mayers:2015} model. For ML TMDs in an dielectric environment, for instance, on a substrate, however exciton calculations become intractable with a standard GW plus BSE approach and have resort to simplified models such as the effective mass model incorporating a Keldysh {\it e-h} interaction  \cite{Berkelbach:2013,Kylanpaa:2015,Mayers:2015} or other effective {\it e-h} interactions \cite{Defo:2016}. While most calculations focus on the exciton ground states of ML TMDs there is a lack of calculations of their exciton spectrum \cite{Qiudy:2013,Yez:2014}. Recently Olsen {\it et al.} have improved the 2DHM and proposed a simple screened hydrogen model (SHM) with an analytical expression for the Rydberg series \cite{Olsen:2016},    
in which an effective dielectric constant dependent on the excitonic level is defined by averaging the above dielectric function $\varepsilon(\mathbf{q})$, linear in $q$,  over the extent of the exciton.  The SHM has reproduced the nonhydrogenic Rydberg series for the $s$ excitons, i.e., orbital quantum number $m=0$, in a freestanding ML of WS$_2$ \cite{Olsen:2016}, but 
it is unclear as yet whether it can make an accurate description for the entire exciton spectrum including a number of $m$ values.

In this paper, we study 2D exciton spectra of a TMD monolayer surrounded by various dielectric environments. We calculate exciton states numerically, using the effective mass model for the excitonic Hamiltonian while employing the Keldysh potential to describe the screened {\it e-h} interaction.  The original SHM deals with a freestanding monolayer alone and we extend it to include screening from the dielectric environment. The SHM is convenient for 2D exciton evaluation, and one of course wants to know the discrepancy between exciton spectra calculated with this model and a more accurate approach. As the the Keldysh potential we employed is closely related to the dielectric function $\varepsilon(\mathbf{q})$ used in the SHM [refer to expression (\ref{vq2d}) in Sec. II below], this allows us to examine the SHM against our numerical results. We found that the SHM can  describe the nonhydrogenic exciton Rydberg series (i.e., the energy's dependence on $n$) reasonably well but it cannot properly describe the variation of the exciton energy with the orbital quantum number $m$. We compare our calculated exciton spectra and binding energies with experimental data for ML TMDs on various substrates. In addition, based on our numerical calculations, we obtain variational wave-functions and analytical expressions for the energy expectation values to easily calculate several strongly bound exciton states. 
Further we use these variational wave-functions to study the Stark effects in ML TMDs, deducing 
an analytical expression that yields a redshift of the ground state energy to a good  accuracy.

This paper is organized as follows. In Sec. II, a formulation of 2D excitons in the effective mass model  is presented where the 2D excitonic equation is simplified to a one-dimensional (1D) differential equation, and the boundary conditions are derived by employing the asymptotic properties of the Keldysh potential. Then an shooting method including Runge-Kutta integration is developed for  the numerical solution of the 1D radial equation (RE). Our extension of the SHM to account for screening from the dielectric environment is also described. In Sec. III, first we present the results of the exciton spectrum of a freestanding ML of MoS$_2$ from the numerical solution of the 1D RE, and then compare with the exciton spectrum from the SHM. We then compare several strongly bound exciton states of ML MoS$_2$ on various substrates calculated with these two approaches (i.e., 1D RE and SHM) as well as our variational method. 
Then, we show results of the 2D excitons in ML WS$_2$ on the SiO$_2$ substrate calculated with the three approaches and comparisons with the experimental data, followed by a comparison of our ground state exciton energies, for the monolayer TMDs in various dielectric environments, with other calculations and also experiments.  Further we present Stark effects in ML TMDs such as the energy shifts and level splitting obtained with our variational wave-functions.   Finally, Sec. IV summarizes the main results obtained. 

\section{Model}
 
\subsection{Excitonic radial equation}
 
Within the effective mass model, the excitonic Hamiltonian can be written as  \cite{Berkelbach:2013,Kylanpaa:2015,Mayers:2015,Robertc:2018,Defo:2016} 
 \begin{equation}
H= - \frac{\hbar^2}{2\mu }\nabla _{\mathbf{r}}^2-V(\mathbf{r}), 
\label{hha0}
 \end{equation}
where $\mathbf{r}=(x,y)$ is the position vector in the plane of the monolayer, and $\mu$ is the exciton reduced mass,  $1/\mu=1/m_e+1/m_h$, $m_e$ and $m_h$ being the electron and hole effective masses. 
As a large spin splitting of the valence bands ($\sim$ 0.15-0.5 eV) was predicted for monolayer TMDs \cite{Xiao:2012} and also measured in optical absorption spectra \cite{Mak:2013, Klots:2014}, we neglect 
mixing of interband transitions associated with excitons A and B, and consider only the A excitons, corresponding to the energy range of 1.8-2.0 eV of typical excitonic absorption spectra. This simpler approach has been  used in previous studies \cite{Berkelbach:2013,Andersen:2015,Kylanpaa:2015,Mayers:2015,Robertc:2018,Olsen:2016,Defo:2016}, and yielded exciton energy levels in agreement with experiments \cite{Chernikova1:2014,Hill:2015}. 
  
For a TMD monolayer surrounded by media with dielectric constants $\varepsilon_a$ (above) and $\varepsilon_s$ (below), 
the effective 2D interaction can be described by the Keldysh potential \cite{Keldysh:1978,Cudazzo:2011,Kylanpaa:2015,Pedersen:2016,Scharf:2016}, 
\begin{equation}
V(r)= \frac{\pi e^2}{2r_0}\left[ H_0\left( \frac{\varepsilon r}{r_0} \right) - Y_0\left( \frac{\varepsilon r}{r_0} \right) \right],  
\label{vkeld}
\end{equation}
where ${H_0}$ and ${Y_0}$ are the Struve function and the Bessel function of the second kind. The length ${r_0}$ relates to the 2D polarizability $\alpha_{2D}$ of the planar material, $r_0=2\pi\alpha_{2D}$, and $\varepsilon$ is the average dielectric constant of the environment, given by $\varepsilon=(\varepsilon_a+\varepsilon_s)/2$.

As $H$ commutes with $L z$, the projection of the orbital angular momentum on the $z$ axis, $[H,L_z]=0$, the orbital
angular momentum along the $z$ axis is conserved, and 
$H$ and $L z$ have simultaneous eigenstates. As $L_z$'s eigenfunctions are $e^{im\theta}$, $m$ being the orbital quantum number, 
we write the eigenfunctions of the Hamiltonian $H$ in a general form as 
\begin{equation}
\psi(\mathbf{r})=\frac{1}{\sqrt{2\pi}}R(r)e^{im\theta} \quad  m= 0, \pm1, \pm2, \ldots, 
\label{psi1}
\end{equation}
which are the solutions to 
\begin{equation}
H\psi(\mathbf{r})=E\psi(\mathbf{r}). 
\label{eigpsi}
\end{equation}
Inserting $H$ and the exciton wave function $\psi(\mathbf{r})$ into Eq.~(\ref{eigpsi}), we find the differential equation for the radial function $R(r)$, 
\begin{equation}
\frac{d^2R}{dr^2}+\frac{1}{r}\frac{dR}{dr}+\frac{2\mu}{\hbar^2}V(r)R -\frac{m^2}{r^2}R=-\frac{2\mu}{\hbar^2}ER~.  
\label{eigR}
\end{equation}
Solving the radial equation (\ref{eigR}) yields eigenenergy $E$ and radial function $R$. Quantization from Eq.~(\ref{eigR}) introduces additional quantum number $n$, namely, the principal quantum number, and thus the complete form of the exciton energy $E$ and wave function $\psi(\mathbf{r})$ can be written as $E_{nm}$ and $\psi_{nm}(\mathbf{r})=R_{nm}(r)e^{im\theta}/\sqrt{2\pi}$, respectively. Given $n$ (n=1, 2, 3,...), then $\lvert m\rvert=0, 1, 2,..., n-1$ \cite{Yang:1991}.  Further, as $m$ enters the RE via the term $-m^2R/r^2$, $E_{nm}$ and $R_{nm}(r)$ depend on only the absolute value of $m$, that is, 
$E_{nm}=E_{n\lvert m\rvert}$ and $R_{nm}(r)=R_{n\lvert m\rvert}(r)$, making the energy levels associated with $\pm m$ doubly degenerate for a nonzero $m$. 

The asymptotic properties of the Bessel and Struve functions result in the 2D interaction having the following asymptotic forms: 
\begin{subequations} 
\begin{equation}
V(r)=-\frac{e^2}{\varepsilon r_0}\left[\ln(\frac{r}{2r_0})+\gamma\right],  \quad \mathrm{for} \quad  r\ll r_0,  
\label{vsmr}
\end{equation}
\begin{equation}
V(r)=\frac{e^2}{\varepsilon r},   \quad \mathrm{for} \quad  r\gg r_0,  
\label{vbgr}
\end{equation}
\end{subequations} 
where $\gamma$ is the Euler-Mascheroni constant, $\gamma=0.57721566\ldots$. 
The above expressions show that the effective 2D potential has a logarithmic divergence at very small distances while it becomes the unscreened Coulomb potential for sufficiently large distances. When $\alpha_{2D}\rightarrow 0$ the wave functions $\psi$ reduce to those of a 2DHM with the Coulomb potential \cite{Yang:1991},
\begin{equation}
V_c(r)=\frac{e^2}{\varepsilon r},  
\label{vcoul}
\end{equation}
corresponding to the energies of the 2D hydrogenic Rydberg series,
\begin{equation}
E_n=-\frac{\mu e^4}{2\varepsilon^2\hbar^2}\frac{1}{(n-1/2)^2}, \quad  n= 1, 2, 3, \ldots.    
\label{E2dcoul}
\end{equation}

The Keldysh and Coulomb potentials [Eqs.~(\ref{vkeld}) and (\ref{vcoul})] are also related through their Fourier transforms. 
Writing the Fourier transform of $V_c(r)$ as $V_{c,\mathbf{q}}=2\pi e^2/(A\varepsilon q)$, with $A$ being the sample area, then the Fourier transform of the Keldysh potential is given by 
\begin{equation}
V_{\mathbf{q}}=V_{c,\mathbf{q}}/(1+\frac{r_0}{\varepsilon}q)=\frac{2\pi e^2}{Aq(\varepsilon+r_0q)}.   
\label{vq2d}
\end{equation}
Therefore the total wavevector-dependent dielectric function, including contributions of the monolayer and its surrounding media, has the form,  
\begin{equation}
\varepsilon(\mathbf{q})=\varepsilon+r_0 q. 
\label{epsdfq}
\end{equation}

The RE (\ref{eigR}) can be transformed as 
\begin{equation}
r\frac{d}{dr}r\frac{dR}{dr}-m^2R+\frac{2\mu}{\hbar^2}\left(Er^2-V(r)r^2\right)R=0.  
\label{eqRr1}
\end{equation}
 
For a small $r$, when the asymptotic expression (\ref{vsmr}) is used for $V(r)$, one finds $\lim_{r\rightarrow 0}r^2V(r)$=0, and the radial equation (\ref{eqRr1}) reduces to 
\begin{equation}
r\frac{d}{dr}r\frac{dR}{dr}-m^2R=0, 
\label{eqRr1b}
\end{equation}
yielding 
\begin{equation}
R(r)=Br^{\lvert m\rvert},  \quad \mathrm{for} \quad  r\rightarrow 0,  
\label{Rrat0}
\end{equation}
where $B$ is a constant, as this form of $R$ ensures that the wave function is finite at $r=0$ \cite{Landau:1977}. 
We note that the radial function has the same asymptotic form at the origin as that for a 2D hydrogen atom. 
From a general consideration of the potential energy $-V(r)$ it is evident that the negative eigenenergies form a discrete spectrum, while the positive energies lie in the continuous spectrum. For large $r$, we neglect the terms in $1/r$, $1/r^2$  and $V(r)$ [as $V(r)\propto 1/r$, refer to Eq.~(\ref{vbgr})] of Eq.~(\ref{eigR}) \cite{Landau:1977} and obtain 
\begin{equation}
\frac{d^2R}{dr^2}=-\frac{2\mu}{\hbar^2}ER~.  
\label{eqRra}
\end{equation}
Then we find the asymptotic behaviour of the radial function $R$ for large $r$, 
\begin{equation}
R(r)\propto e^{-\sqrt{-2\mu E}r/\hbar},  \quad \mathrm{for} \quad  r\rightarrow \infty,  
\label{Rrat9}
\end{equation}
which vanishes at infinity. 

Introduce $u$ by the substitution $R=u/r$ and scale $r$ by a factor $1/r_0$, $\rho=r/r_0$. Then we nondimensionalize Eq.~(\ref{eigR}) and transform it after substituting expression (\ref{vkeld}) for $V(r)$ into the following differential equation, 
\begin{equation}
u^{\prime\prime}-\frac{1}{\rho}u'+ \left[\gamma\left(H_0(\rho) - Y_0(\rho)\right) +\frac{1-m^2}{\rho^2}-\lambda \right]u=0, 
\label{eigu}
\end{equation}
where the primes denote derivatives with respect to $\rho$, $\gamma=\mu\pi e^2r_0/(\varepsilon\hbar^2)$, and the eigenvalue $\lambda$ relates to the exciton energy $E$ via $\lambda=-2\mu r_0^2E/\hbar^2$, both $\gamma$ and $\lambda$ being dimensionless. 

To solve Eq.~(\ref{eigu}) we need the boundary conditions.  From the asymptotic form of Eq.~(\ref{Rrat0}) for $R$ in the neighbourhood of $r=0$ one finds
\begin{equation}
u(\rho)=C\rho^{\lvert m\rvert+1},  \quad \mathrm{for} \quad  \rho\rightarrow 0, 
\label{uat0}
\end{equation}
where $C$ is a constant, showing $\lim_{\rho\rightarrow0}u(\rho)=0$.

For large distances the boundary conditions are determined by $R$'s asymptotic form Eq.~(\ref{Rrat9}) and are given by 
\begin{equation}
\lim_{\rho\rightarrow\infty}u(\rho)=0,  \quad  \lim_{\rho\rightarrow\infty}u'(\rho)=0. 
\label{uat9}
\end{equation}
Eq.~(\ref{eigR}) describes one-dimensional motion in a semi-infinite region $(0,\infty)$. For a given $\lvert m\rvert$, none of the energy levels is degenerate;  there is only one eigenfunction $R(r)$ corresponding to the eigenenergy $E$. 

To find eigensolutions the boundary value problem [Eq.~(\ref{eigu})] is numerically solved using a    shooting method in conjunction with fourth-order Runge-Kutta integration.  A very small $\rho$, denoted as $\rho_0$,  and a sufficiently large $\rho$, denoted by $\rho_x$, are taken such that boundary condition (\ref{uat0}) is used at $\rho_0$, $u(\rho_0)=C\rho_0^{\lvert m\rvert+1}$, while the boundary conditions (\ref{uat9}) are used for the $\rho_x$ point, $u(\rho_x)=0$, $u'(\rho_x)=0$. The length $\rho_x-\rho_0$ is 
discretized into slices of interval $\Delta\rho$. Beginning with the first slice, $\rho\in[\rho_x-\Delta\rho,\rho_x]$, for an initial value of $\lambda$, integrate inward by a Runge-Kutta method from $\rho_x$ to $\rho_x-\Delta\rho$ to find the values for $u(\rho_x-\Delta\rho)$, $u'(\rho_x-\Delta\rho)$. Iterating integration over the rest of the intervals yields a trajectory $u(\rho;\lambda)$ with  the endpoint value of $u(\rho_0;\lambda)$. If $u(\rho_0;\lambda)\neq u(\rho_0)$, then by varying the $\lambda$ value we produce trajectories until we find the trajectory which has the desired boundary value at endpoint $\rho_0$, and the immediate $\lambda$ value corresponding to this trajectory is just the eigenvalue (that is related to exciton energy $E$) we seek for Eq.~(\ref{eigu}). The interval $\Delta\rho$ is taken to be 0.0001. $\rho_0$ varies according to $m$, while  
the $\rho_x$ value varies depending on the upper bound set for the discrete exciton energies. For exciton energies lower than $-0.05$ eV, for instance, $\rho_x=25.0$ is used. Tolerances of $10^{-12}$ to $10^{-11}$ are used for $\lambda$ to obtain accurate eigensolutions. 
The radial function $R$ needs to be normalized and the normalizing constant is given by $C_N=(\int_0^{\infty}R^2rdr)^{-1/2}=(\int_0^{\infty}d\rho{u^2/\rho})^{-1/2}$. In the 1D calculation, the above numerical integration is carried out over the $\rho$-mesh by the trapezoidal rule.

 \subsection{Screened hydrogen model}

The original SHM deals with a freestanding ML \cite{Olsen:2016}, and for a ML material on a substrate or encapsulated in a dielectric we can extend the model to include the effects of the dielectric environment by employing the dielectric function $\varepsilon(q)$  [expression~(\ref{epsdfq})].   
An effective dielectric constant $\varepsilon_{eff}$ is defined by averaging $\varepsilon(q)$  in wave-vector space over a disc with radius $1/a_{eff}$, $a_{eff}$ being an effective radius of the exciton, 
\begin{equation}
\varepsilon_{eff}=\varepsilon+\frac{4\pi}{3} \alpha_{2D}/a_{eff}.    
\label{epseff11}
\end{equation}
The exciton Bohr radius $a_{nm}=\langle \psi_{nm} \vert r\vert \psi_{nm} \rangle=\int R_{nm}^2r^2dr$ is taken as $a_{eff}$ ($\psi_{nm}$ is the exciton wave-function), $a_{eff}=a_{nm}$, such that the effective dielectric constant $\varepsilon_{eff}$ becomes dependent on $n,m$, the exciton state,  that is, $\varepsilon_{eff}=\varepsilon_{nm}$.  
The exciton Bohr radius $a_{nm}$ of the SHM \cite{Olsen:2016} is assumed to equal the state-dependent Bohr radius of the 2D hydrogen atom  \cite{Yang:1991} multiplied by a factor of $\varepsilon_{nm}$, 
\begin{equation}
a_{nm}=\varepsilon_{nm}\hbar^2[3n(n-1)-m^2+1]/(2\mu e^2).   
\label{anmeff}
\end{equation}
Combing Eqs.~(\ref{epseff11}) and (\ref{anmeff}) then one finds $\varepsilon_{nm}$, 
\begin{equation}
\varepsilon_{nm}=\frac{1}{2}\varepsilon\left\{1+\sqrt{1+\frac{32\pi\alpha_{2D}\mu e^2}{3\varepsilon^2\hbar^2[3n(n-1)-m^2+1]}}\right\}. 
\label{epseff12}
\end{equation}
Substituting the above expression for $\varepsilon_{nm}$ in Eq.~(\ref{anmeff}) gives the explicit expression for the exciton effective  radius of the SHM. 
The exciton spectrum is assumed to have the form of expression (\ref{E2dcoul}) for the 2D hydrogenic Rydberg series, with  
only $\varepsilon$ there replaced by the state-dependent effective dielectric constant $\varepsilon_{nm}$, 
\begin{equation}
E_{nm}=-\frac{\mu e^4}{2\varepsilon_{nm}^2\hbar^2}\frac{1}{(n-1/2)^2}.   
\label{exoneff}
\end{equation}

In this study, two key parameters of monolayer TMDs, the exciton reduced mass $\mu$ and the 2D polarizability $\alpha_{2D}$, are needed, which are taken from Refs. \cite{Berkelbach:2013,Olsen:2016}, obtained from first-principles calculations in density functional theory (DFT).

\section{Results and discussions}

\subsection{Exciton spectra: RE solution versus SHM}

We begin with a freestanding monolayer of MoS$_2$ ($\varepsilon_a=\varepsilon_s=1$), corresponding to the suspended monolayer samples in experiments \cite{Klots:2014}.  The electron and hole effective masses are taken to be 0.5 $m_0$ ($m_0$ is the electron rest mass), giving the exciton reduced mass $\mu=0.25m_0$. We use the 2D polarizability  6.6 $\AA$ \cite{Berkelbach:2013},  corresponding to a screening length $r_0=41.5 \AA$ . 

Figure~\ref{fig1} shows eight Rydberg series of exciton energy levels for $E_{n\lvert m\rvert}$ up to -0.04 eV, corresponding to eight orbital quantum number $\lvert m\rvert=0-7$, calculated from the RE. The ground state, $n=1$, $m=0$,  which is denoted by $1s$ following conventional notation for the 2D hydrogen atom \cite{Yang:1991,He:2014,Yez:2014,Olsen:2016}, has a binding energy
0.555 eV that is very close to the value 0.54 eV calculated using the same parameters \cite{Berkelbach:2013}. Above the $1s$ state is the doubly degenerate $2p$ exciton states of $n=2$, $m=\pm1$, followed by the $2s$ state, $n=2$, $m=0$. 
We also see quasicontinuum states above a series of discrete exciton levels. Compared to the 2D hydrogenic Rydberg series [expression (\ref{E2dcoul})], the exciton energy levels are elevated having smaller magnitudes; for the ground state, for instance, the 2D hydrogen model yields an energy of -13.6 eV with reduced mass $\mu=m_0/4$, far lower than the -0.555 eV energy. Neglecting screening due to the monolayer ($\alpha_{2D}=0$) the Keldysh interaction  [Eq.~(\ref{vkeld})] becomes simply the 2D Coulomb potential [Eq.~(\ref{vcoul})]. The difference between the exciton spectrum [Fig.~\ref{fig1}] and the hydrogenic Rydberg series is entirely due to the 2D dielectric screening, which has weakened the electron-hole Coulomb attraction and thus reduced the exciton binding energies while raising the exciton energies.   

With an {\it e-h} Coulomb interaction, as expression (\ref{E2dcoul}) shows, all exciton states with different $m$ but the same $n$ have the same energy (i.e., the Coulomb degeneracy); now we see from Fig.~\ref{fig1} that this degeneracy of exciton levels is lifted as the Keldysh interaction in ML MoS$_2$ deviates significantly from the Coulomb potential. For the same $n$, the principal quantum number, the energy level of $E_{n\lvert m\rvert}$ is lower for a larger $\lvert m\rvert$, a phenomenon that has been found in a previous study using a first-principles GW-BSE approach \cite{Yez:2014}.  
This result indicates that the 2D dielectric screening associated with exciton ($n$,$m$) becomes weaker as $\lvert m\rvert$ increases (for a fixed $n$). To explain this we look at the effective Bohr radius of exciton ($n$,$m$).   
The effective Bohr radii of six strongly bound exciton states are given in Fig.~\ref{fig1}, bracketed just below the exciton energy levels (values in $\AA$). The 2D excitons are of Wannier type as their effective radii are much larger than the unit cell dimensions (the lattice constant of monolayer MoS$_2$ is 3.16 $\AA$). 
Our calculation shows that the exciton Bohr orbit shrinks as $\vert m \vert$ gets larger for the same $n$.  
A Bohr orbit of shorter radius $a_{nm}$ corresponds to a stronger electron-hole interaction and weaker dielectric screening, according to Eq.~(\ref{vkeld}), thus giving rise to a larger exciton binding energy.  Furthermore, 
as angular momentum $L_z$ is conserved, its eigenvalue being $m\hbar$ in exciton state $\psi_{nm}$, we write 
$\lvert m\rvert\hbar=\mu\omega_{nm} a_{nm}^2$, where  $\omega_{nm}$ is the orbital frequency (circular). 
 Evidently  
a greater angular momentum $\lvert m\rvert\hbar$ together with a shorter exciton Bohr radius $a_{nm}$ leads to a larger frequency $\omega_{nm}$ or shorter orbital period for the exciton, a further result from the reduced dielectric screening. 

The energy levels get closer to those of the 2DHM [Eq.~(\ref{E2dcoul})] as $n$ increases [for instance, for the $8s$ exciton $n=8$, $m=0$, its energy -0.037 eV, is only 0.02 eV higher than the energy $E_8$ given by Eq.~(\ref{E2dcoul})], because at a larger distance $r$ (or equally, a larger exciton Bohr radius $a_{nm}$) the Keldysh interaction approaches more closely the 2D Coulomb  potential.  
The number of nodes of the radial functions $R_{nm}(r)$ follows that of a 2D hydrogen atom, $n-\lvert m\rvert-1$, as is illustrated in Fig.~\ref{fig2} for the $R_{nm}(r)$ of the six low-energy levels, $1s$, $2s$, $2p$, $3s$, $3p$, $3d$.

In Figs.~\ref{fig3}(a), \ref{fig3}(b) and \ref{fig3}(c) we show the exciton Rydberg series by the absolute values of the energies for $\lvert m\rvert=0, 1, 2$, respectively, calculated with the SHM [Eqs.~(\ref{epseff12}) and (\ref{exoneff})] and from the RE (\ref{eigR}).  We see that the exciton energies evaluated with the SHM are in good agreement with our numerical calculation except that the binding energies of  the $1s$, $2p$, $3d$ excitons, i.e., the most strongly bound excitons, are underestimated by  23\%, 32\%, 30\%, respectively.  Such a discrepancy in exciton energy can be explained as follows. In the SHM, the averaging over $q$ to obtain the effective screening $\varepsilon_{eff}$ [expression~(\ref{epseff11})] is carried out for only long wavelengths, $\lambda >a_{eff}$, and the shorter wavelengths (larger-$q$) contribution may become appreciable for excitons with a small radius or equally a large binding energy. 
In addition, the energy expression~(\ref{exoneff}) has a form taking from the 2D hydrogenic Rydberg series [expression (\ref{E2dcoul})], a result from the 2D Coulomb potential, and is only a good approach when dealing with excited states with a large exciton radius. 
The exciton energy discrepancies above correspond to the overestimates of the 2D dielectric screening by 14\%, 21\%, 20\% for excitons $1s$, $2p$, $3d$ respectively, which also explains  
 the sharp rise of the effective dielectric constant occurring on the small $n$ side for $\lvert m\rvert=0, 1, 2$,
as shown in the insets of Fig.~\ref{fig3}.  
An underestimate with the SHM of the ground state exciton binding energies of monolayer TMDs has also been noted in a previous study \cite{Olsen:2016}. 
The exciton effective radius $a_{nm}$ is a key parameter of the SHM, and $a_{nm}$ versus $n$ for $\lvert m\rvert=0, 1, 2$ are shown in Fig.~\ref{fig4}(b); comparing with the RE result [Figs.~\ref{fig4}(a)] we see that the SHM has captured the main character of the radius's dependence  on $n$.  

Now we look at how the energies of the exciton spectrum vary quantitatively with the orbital quantum number $m$. 
Our RE solution shows that for a given $n$, the principle quantum number, the exciton binding energy $\lvert E_{nm}\rvert$ increases with the orbital  quantum number $m$ [Fig.~\ref{fig5}(a)], and further the increase is linear, with a slope that decreases as $n$ becomes larger. In contrast, the SHM yields very different results [Fig.~\ref{fig5}(b)], which predicts that the energy $\lvert E_{nm}\rvert$ decreases as $m$ increases, in a nonlinear manner.  
Further, the exciton effective radius $a_{nm}$ of the SHM varies with $m$ in a manner [Fig.~\ref{fig6}(b)] that is very different from the result calculated from the RE in Fig.~\ref{fig6}(a), which shows a {\it linear} $m$-dependence with the slope remaining almost  unchanged as $n$ varies.  In fact, the SHM's effective radius varies with $m$ in a manner very similar to that of the 2DHM [Fig.~\ref{fig6}(c)] containing a {\it quadratic}  $m^2$  dependence [refer to expression~(\ref{anmeff})], the difference being that the SHM yields a slightly larger effective radius after accounting for the dielectric screening [comparing Figs.~\ref{fig6}(b) and \ref{fig6}(c)].  
As a result, the SHM has failed to describe the variation of the exciton energy $E_{nm}$ with orbital quantum number $m$ of the exciton spectrum.

\subsection{2D excitonic variational wave functions}

The excitons in quantum wells can be described analytically with the 2DHM \cite{Haug:2004,Parfitt:2002}, in which the wave function of the ground state is $\varphi_{10}(\mathbf{r})=\frac{4}{\sqrt{2\pi} a_0} e^{-2r/a_0}$, and the first excited states are triply degenerate, 
$\varphi_{20}(\mathbf{r})=\frac{4}{3\sqrt{6\pi} a_0}(1-\frac{4r}{3a_0}) e^{-2r/3a_0}$,
$\varphi_{2\pm1}(\mathbf{r})=\frac{8}{9\sqrt{3\pi} a_0}\frac{r}{a_0}e^{-2r/3a_0}e^{ \pm i\theta}$, where $a_0$ is the exciton Bohr  radius, $a_0=\hbar^2\varepsilon/(\mu e^2)$, $\varepsilon$ being the dielectric constant of the quantum well material.  
For excitons in monolayer TMDs that have an effective interaction  as given by expression~(\ref{vkeld}), analytical expressions will be very useful for practical calculations of the 2D exciton properties such as the binding energies, internal exciton transitions (e.g., $1s-2p$ or  $2s-2p$ transition induced by an external electric field \cite{Dent:2003}) and Stark effects (see subsection D below). Considering that the wave functions have a nodal structure (Fig.~\ref{fig2}) similar to that of the 2D hydrogenic wave-functions, we can work out variational wave functions $\phi_{nm}(\mathbf{r})$ in the following forms for the first three energy levels, 
\begin{subequations} 
\begin{equation}
\phi_{10}(\mathbf{r})=\sqrt{\frac{2}{\pi a^2}} e^{-r/a}, 
\label{fi00}
\end{equation}
\begin{equation}
\phi_{2\pm1}(\mathbf{r})=\frac{2r}{\sqrt{3\pi}a'^2}e^{-r/a^\prime}e^{ \pm i\theta }, 
\label{fi11}
\end{equation}
\begin{equation}
\phi_{20}(\mathbf{r})=\frac{4}{\sqrt{3\pi\eta}a''}\left[\frac{1}{2}\left(1+\frac{a''}{a}\right)\frac{r}{a''}-1\right]e^{-r/a''},  
\label{fi10}
\end{equation}
\end{subequations} 
where $a$, $a'$, $a''$ are the variational parameters, and $\eta$ is dimensionless,  introduced for normalization, $\eta=(a''/a)^2- \frac{2}{3}a''/a+1$. These trial wave functions $\phi_{nm}(\mathbf{r})$ have similar forms to the hydrogenic wave functions $\varphi_{nm}(\mathbf{r})$ above, and further they are orthogonal and normalized. 

To calculate the expectation values of $H$, namely, $E_{10}(a)=\langle \phi_{10} \vert H \vert \phi_{10} \rangle$,  $E_{2\pm1}(a')=\langle \phi_{2\pm1} \vert H \vert \phi_{2\pm1} \rangle$, $E_{20}(a'',a)=\langle \phi_{20} \vert H \vert \phi_{20} \rangle$, which involve integrals over $\mathbf{r}=(r,\theta)$, we expand the Keldysh potential $V(r)$ [Eq.~(\ref{vkeld})], $V(r)=\sum_{\mathbf{q}}V_{\mathbf{q}}e^{i\mathbf{q}\cdot\mathbf{r}}$, where $V_{\mathbf{q}}$ is given by Eq.~(\ref{vq2d}), such that the integration over $\mathbf{r}$ can be performed analytically, and further the integration over $q,\varphi$ is also obtained analytically after converting the summation over $\mathbf{q}$ to a double  integral. After a lengthy derivation then we obtain for the expectation values the following expressions,  
\begin{subequations} 
\begin{equation}
E_{10}(a)=\frac{\hbar^2}{2\mu a^2} - \frac{2e^2}{\varepsilon a}P_0(\frac{2r_0}{\varepsilon a}),  
\label{E10}
\end{equation}
\begin{equation}
E_{2\pm1}(a')=\frac{\hbar ^2}{2\mu a'^2} + \frac{e^2}{\varepsilon a'}\left[3P_1(\frac{2r_0}{\varepsilon a'}) - 5P_2(\frac{2r_0}{\varepsilon a'})\right], 
\label{E21}
\end{equation}
\begin{align}
E_{20}(a'',a)&= \frac{8}{\eta}\left\{\Big(\nu^2+\frac{1}{2}\Big)\frac{\hbar^2}{3\mu a''^2}
+\frac{2e^2}{3\varepsilon a''} \Big[ 
-(2\nu+1)P_0(\frac{2r_0}{\varepsilon a''}) \Big. \right.
\nonumber \\ 
&\qquad {}  \left. \Big. +3\nu (3\nu+2) P_1(\frac{2r_0}{\varepsilon a''})-15\nu^2  P_2(\frac{2r_0}{\varepsilon a''}) \Big]\right\},   
\label{E20}
\end{align}
\end{subequations} 
where $\nu=\frac{1}{4}(1+a''/a)$, and 
the dimensionless functions $P_0(x)$, $P_1(x)$, $P_2(x)$ are given by  
\begin{subequations} 
\begin{equation}
P_0(x)=\frac{x^2}{1+x^2}\left\{\frac{1-x}{x^2}+(1+x^2)^{-1/2}\left[\sinh^{-1}(x)+\sinh^{-1}(\frac{1}{x})\right]\right\},  
\end{equation}
\begin{equation}
P_1(x)=\frac{x^2}{1+x^2}\left[\frac{2-x}{3x^2}+P_0(x)\right], 
\end{equation}
\begin{equation}
P_2(x)=\frac{x^2}{1+x^2}\left[\frac{8-3x}{15x^2}+P_1(x)\right],  
\end{equation}
\end{subequations} 
$\sinh^{-1}(x)$ being the inverse hyperbolic sine, $\sinh^{-1}(x)=\ln(x+\sqrt{x^2+1})$. We have checked and verified these expressions by comparing their values with those $H$'s expectation values obtained by numerically integrating $\langle \psi_{nm} \vert V \vert \psi_{nm} \rangle$, with expression (\ref{vkeld}) for $V(r)$, using a Gauss-Legendre quadrature method. 
The energy expressions for the higher-energy levels such as $E_{20}$ become more complicated as more variational parameters are needed in the orthogonalization of the wave functions.

Continuing freestanding monolayer MoS$_2$, 
we plot the expectation values of $E_{10}$ and $E_{2\pm1}$ as functions of variational parameters $a$ and $a'$, respectively, in Figs.~\ref{fig7}(a) and \ref{fig7}(b). From these  
one finds the minimum of $E_{10}$ at $a$ = 10.4 $\AA$ and the minimum of $E_{2\pm1}$ at $a'$ = 11.2 $\AA$, yielding the $1s$ and $2p$ exciton energies $E_{10}$=-0.543 eV, and  
$E_{2\pm1}$=-0.312 eV. Having the $a$ value, one then plots the $E_{20}$ versus $a''$ curve [Fig.~\ref{fig7}(c)] that gives the $2s$ exciton energy $E_{20}$=-0.242 eV with  
$a''$=16.1 $\AA$. These exciton energies are very close to the RE solution above, the former being larger by 2\%, 2\%, 6\% for the $1s$, $2p$, $2s$ exciton, respectively. 
Furthermore, the wave-functions obtained from the variational method and RE solution are very close, as shown in Fig.~\ref{fig8}. 

We calculated exciton spectra of monolayer MoS$_2$ on various substrates by solving 
the RE (\ref{eigR}), with average background dielectric constant $\varepsilon=(1+\varepsilon_s)/2$.   To further check the variational method (VM)  [Eqs.~(\ref{E10})-(\ref{E20})] and SHM in strongly bound exciton calculation, we compared the results of the two methods with the RE calculation for the $1s$, $2p$, $2s$ excitons (Table~\ref{table:1}). The SHM yields a $1s$ exciton binding energy $\lvert E_{10}\rvert$ which is 20\% lower for a freestanding monolayer and 10\% lower for monolayer MoS$_2$ on substrate SiO$_2$ (with a smaller $\varepsilon_s$), but $\sim$9\% higher for  monolayer MoS$_2$ on substrate hBN or diamond (with a larger $\varepsilon_s$). For the $2s$ and $2p$ excitons, the SHM predicts $\lvert E_{20}\rvert >\lvert E_{21}\rvert$, which is different from the RE result,  $\lvert E_{20}\rvert <\lvert E_{21}\rvert$, consistent with the above finding from Fig.~\ref{fig5}.  The VM and RE solution again yield very close exciton energies, with the $2s$ energies having the largest deviation of 6\%.

\subsection{Comparisons with experiment}

Figure~\ref{fig9}(a) shows the exciton spectrum $E_{nm}$ of monolayer WS$_2$ on the SiO$_2$ substrate ($\varepsilon_s$=2.1, see the Supplemental Material of Ref. \cite{Chernikova1:2014}) consisting of the $s$ ($m$=0), $p$ ($m$=1), $d$ ($m$=2) excitons, calculated from the RE using the reduced mass and 2D polarizability $\mu=0.22m_0$ and $\alpha_{2D}$=6.35 $\AA$ as obtained from DFT calculations in the Supplemental Material of Ref.\cite{Olsen:2016}. The $1s$, $2p$, $2s$ excitons remain deeply confined, and their energies are very close to those of the variational calculation [solid squares in Fig.~\ref{fig9}(a)].  The binding energies are significantly reduced due to the {\it additional} screening from the substrate, and the excitons with energies above the $5s$ state approach the quasi-continuum states. We see again that for the same principle quantum number $n$ excitons with a larger orbital quantum number $m$ have a lower energy and accordingly a larger binding energy, for instance, $E_{32}<E_{31}<E_{30}$. For a given $n$, the exciton energy $E_{nm}$ decreases linearly with $m$ (shown in Fig.~\ref{fig10}), similar to the  freestanding monolayer case. The $s$ exciton states were probed in measurements of the linear optical spectra of monolayer WS$_2$ on a SiO$_2$ substrate \cite{Chernikova1:2014}), their energy levels deviating significantly from the 2D hydrogenic Rydberg series of expression (\ref{E2dcoul}). These experimental data are shown in Fig.~\ref{fig9}(b) (solid circles) to make a quantitative comparison with our RE calculation (diamonds) and also the results of the SHM (triangles) and 2DHM (stars). Our calculated exciton energies are in good agreement with experiment except for the $1s$ energy, which is $\sim$0.09 eV lower than the experimental value, similar to the discrepancy 0.08 eV given in the Supplemental Material of Ref.\cite{Chernikova1:2014}). The SHM also makes a good prediction when neglecting its overestimation of the $1s$ exciton binding energy. In contrast, the 2DHM yields much larger exciton binding energies, in particular for the low-lying exciton states; for instance, the 2DHM ground state binding energy 5 eV is more than one order of magnitude larger than the experimental value 0.32 eV. Unlike the $s$ states, the $p$ states are excitonic dark states as they do not appear in the linear optical spectrum. In another experimental study \cite{Yez:2014}, both $s$ and $p$ exciton energy levels were measured for monolayer WS$_2$ on substrate SiO$_2$, and in particular the $p$ exciton states were probed using two-photon excitation spectroscopy. In the measured spectrum of $p$ excitons (Fig.~2 of Ref.\cite{Yez:2014}), there are two broad features of spectral widths 0.11 and 0.07 eV, respectively, peaking at energies 0.24 and 0.43 eV, respectively, above the $1s$ state. In fact our calculated $1s$-$2p$ separation 0.22 eV is very close to the experimental value of 0.24 eV; further the  $1s$-$3p$ energy interval we calculated, 0.3 eV, suggests that the $3p$ excitons may contribute to the lower-energy absorption feature while broadening its energy range.  The higher-energy feature can be attributed to the absorption due to $4p$, $5p$, $6p$ excitons, which appear energetically 0.37 to 0.41 eV higher than the $1s$ state [Fig.~\ref{fig9}(a)]. 

The ground state exciton energies of freely suspended ML TMDs and ML TMDs on a SiO$_2$ or hBN substrate have been calculated using various approaches, which are  summarized in Table~\ref{table:2} (columns 6 and 7).  Clearly more calculations were performed for suspended monolayers, for which the binding energy values obtained from the GW plus BSE approach vary with a difference that can reach $\sim$0.5 eV.  
We also calculated the ground state exciton binding energies for these TMD monolayers in three different dielectric environments as shown (column 5 of Table~\ref{table:2}),  
using the exciton reduced mass and 2D polarizability values (listed in columns 2 and 3 of Table~\ref{table:2}) from the DFT calculations of Ref.\cite{Olsen:2016}  (Supplemental Material therein). We see that the binding energies become smaller for the monolayers on a substrate with stronger dielectric screening (i.e. greater $\varepsilon_s$), and also the binding energies we calculated are very close to those obtained previously with a similar approach, that is, the effective mass model in conjunction with the Keldysh interaction, but are 10-20\% smaller than the binding energies given in Ref.\cite{Defo:2016}. We note however that these calculated binding energies necessarily depend on the reduced mass, the 2D polarizability and the exact form of the electron-hole interaction potential. Experimentally, ground state exciton binding energies have been measured for isolated MoS$_2$ monolayers and TMD monolayers on a substrate, most on SiO$_2$ or fused silica \cite{Chernikova1:2014,Yez:2014,He:2014,Hill:2015,Rigosi1:2016,Liuhj:2015,Stier:2016,Wangg2:2015,Zhubr1:2015,Huangja1:2016,Diware:2018,Hanbicki:2015}, which are also listed in Table~\ref{table:2} (the last two columns) for a quantitative comparison. We have found no measurement on WSe$_2$ monolayers on a hBN substrate and instead put an experimental value of monolayer WSe$_2$ on diamond which has a similar dielectric constant to hBN. Photocurrent measurements on a suspended monolayer of MoS$_2$ obtained a lower bound for its exciton binding energy 0.57 eV \cite{Klots:2014}; in such    dielectric environments as the dielectric screening ($\varepsilon_s$) increases from vacuum to SiO$_2$ to hBN, the experimental data exhibit a decrease in exciton binding energy, a trend that is in agreement with the theoretical prediction. 
The experimental values fall in the range from 0.3 to 0.7 eV (0.2 to 0.9 eV) for WS$_2$ (WSe$_2$) on the SiO$_2$ substrate, whereas the different models predict a binding energy varying from 0.3 to 0.4 eV approximately, a smaller deviation of 0.1 eV.  
Overall our calculations agree with most of the experimental data.

\subsection{Stark effects}

We now turn to the 2D excitons in an applied in-plane electric field $\mathbf{F}$. Then  the eigenequation is given by $(H+e\mathbf{F}\cdot\mathbf{r})\Psi(\mathbf{r})=E\Psi(\mathbf{r})$, where $H$ is the Hamiltonian in the absence of an electric field [Eq.~(\ref{hha0})].  We confine ourselves to the strongly bound low-energy excitons of $1s$, $2p$, $2s$ (constituting a four-state model system), which dominate the exciton absorption spectrum in the low energy region \cite{Chernikova1:2014,Yez:2014,He:2014,Hill:2015}. The variational wave-functions at zero field [Eqs.~(\ref{fi00})-(\ref{fi10})] that we obtained above can now be used conveniently to study the Stark effects in monolayer TMDs. As the $1s$ and $2s$  wavefunctions are even and the $2p$ wavefunction is odd in parity, the electric field couples the exciton states $1s$ and $2p$ as well as exciton states $2s$ and $2p$, and therefore only matrix elements  $\langle \phi_{2\pm1} \vert e\mathbf{F}\cdot\mathbf{r} \vert \phi_{10} \rangle$ and $\langle \phi_{2\pm1} \vert e\mathbf{F}\cdot\mathbf{r} \vert \phi_{20} \rangle$ are nonzero.  Letting $\mathbf{F}$ along the $x$ axis, one finds
$V_1=\langle \phi_{2\pm1} \vert eFx \vert \phi_{10} \rangle=eFa4\sqrt{6}a^2a'^2/(a+a')^4$, and $V_2=\langle \phi_{2\pm1} \vert eFx \vert \phi_{20} \rangle=eFa''16a'^2a''^2/(a'+a'')^4[2(1+a''/a)a'/(a'+a'')-1]/\sqrt{\eta}$. Given a field strength $F$, the exciton energies are solutions to the secular equation, 
\begin{align}
(E-E_{21})\left\{ E^3-(E_{10}+E_{21}+E_{20})E^2+[E_{10}E_{21}+E_{21}E_{20}+E_{10}E_{20}-2(V_1^2+V_2^2)]E   \right. \nonumber \\ 
 \left. +(2V_2^2E_{10}+2V_1^2E_{20}-E_{10}E_{21}E_{20}) \right\}=0,   
\label{seqfild}
\end{align}
which are shown in Figs.~\ref{fig11}(a) and \ref{fig11}(b) for the ground state ($1s$) and the excited states ($2p$ and $2s$) in freestanding monolayer MoS$_2$.   With an electric field applied, clearly the $1s$ energy level is redshifted while the $2p$ level  splits into two. 
Using second-order perturbation theory, we obtain an analytical expression for the energy shift of the ground state,  
$\delta E_{10}=-\alpha F^2/2$, i.e., the second-order Stark effect, where $\alpha$ is the electric polarizability of the exciton \cite{Pedersen:2016,Scharf:2016}, given by $\alpha=384e^2a^2/(E_{21}-E_{10})(aa')^4/(a+a')^8$. This quadratic dependence of $E_{10}$ on the field $F$ is also plotted in Fig.~\ref{fig11}(a) (dotted curve), showing the perturbative approach yields an accurate correction to the ground state energy for fields below 50 V/$\mu$m. The electric polarizability of the ground state exciton is $\alpha$=7$\times$ 10$^{-18}$ eV(m/V)$^2$; its energy redshift is 1.4 meV at $F$=20 V/$\mu$m, and increases to 9 meV at  
$F$=50 V/$\mu$m, close to the energy shifts of 1.5 and 10 meV for the two field strengths respectively that were reported in Ref.\cite{Scharf:2016}.  
For monolayer MoS$_2$ on the SiO$_2$ substrate (average dielectric constant $\varepsilon$ is taken to be 2.45, the same as in  Ref.\cite{Scharf:2016} for comparison of the exciton energies), the polarizability of the ground state exciton increases to 1.1$\times$ 10$^{-17}$ eV(m/V)$^2$, corresponding to an energy redshift of 2.3 meV at $F$=20 V/$\mu$m, close to the 3 meV redshift in Ref.\cite{Scharf:2016}. For monolayer MoS$_2$ encapsulated in h-BN (average dielectric constant $\varepsilon$=(5+5)/2=5), we find that the ground state energy shift deviates from the quadratic field dependence for fields $F$ exceeding 15 V/$\mu$m (not shown), similar to the finding in Ref.\cite{Scharf:2016};  at low field strengths, we obtain for the ground state exciton a polarizability of  2.1$\times$ 10$^{-17}$ eV(m/V)$^2$, a value smaller than 3.5$\times$ 10$^{-17}$ eV(m/V)$^2$ given in Ref.\cite{Scharf:2016} but larger than 1.4$\times$ 10$^{-17}$ eV(m/V)$^2$ reported in Ref.\cite{Pedersen:2016}. The simpler approach above yields splitting of the $2p$ states, and also offers an accurate description of the ground state and its energy shift. This is because the $1s$ and $2p$ states are strongly bound states and the high energy levels above make a very small contribution. For a more accurate description of the Stark effects, one of course needs to account for these high energy states in further study.

\section{Conclusions}

In conclusion, we have studied 2D exciton spectra of monolayer TMDs using an effective mass model incorporating a screened 2D electron-hole interaction described by the Keldysh potential.  Freestanding monolayer TMDs as well as monolayers on various substrates have been considered. The excitonic Schr{\"o}dinger equation is reduced to a 1D RE, and the boundary conditions for the exciton radial functions are obtained after considering the asymptotic expressions of the Keldysh potential. The exciton energies and wave-functions are numerically calculated by solving the RE with a shooting method including fourth-order Runge-Kutta integration. 
We paid particular attention to the simple models to use for 2D exciton calculation. 
The 2DHM yields much lower exciton energies for ML TMDs, one order of magnitude lower, for instance, for the ground states, than the Rydberg series obtained from the RE. 
We examined the SHM, an improved version of the 2DHM,  which contains an exciton effective radius and an effective dielectric constant, by comparing its exciton spectra with the RE calculations. The SHM described the exciton 
Rydberg series reasonably well. 
 For a given $n$, however the SHM failed to account for the dependence of the exciton energy on the orbital quantum number $m$. The RE results showed that the exciton energy decreases linearly as $m$ increases, and the energy decrease is due to the shrinking exciton Bohr orbit which causes the electron-hole interaction to be enhanced and consequently the exciton binding energy to be raised. The exciton effective radius expression~(\ref{anmeff}), generalized from a 2D hydrogenic result, can characterize  the exciton radius's dependence on $n$, but it cannot properly describe the  exciton radius's dependence on $m$, which is the cause of the SHM's poor description of the $m$-dependence of the exciton energy. 

We also paid attention to two experimental studies on exciton energy levels, one measuring $s$ excitons while the other probing both $s$ and $p$ excitons, for monolayer WS$_2$ on substrate SiO$_2$.  
Our calculated $s$ exciton Rydberg series, which deviates significantly from the 2D hydrogenic Rydberg series, agree well with those measured by optical reflection spectroscopy. 
We also analyzed the two-photon absorption spectrum, and explained its two broad features in terms of the dark $p$ excitons: the lower-energy feature arises due to both $2p$ and $3p$ exciton absorption whereas the higher-energy feature is attributed to the absorption due to the higher energy $p$ excitons. 
Using exciton reduced mass and 2D polarizability values previously calculated in DFT, we calculated exciton energies for monolayer TMDs in various dielectric environments and made comparisons with other numerical calculations and also the experimental data available.  
The exciton binding energies are very close to those calculated with a similar approach and also compare favourably with most of the experimental measurements. A smaller binding energy was predicted for the monolayer in an environment that has stronger dielectric screening, consistent with experimental results. 
Based on the RE calculations and the 2D hydrogenic wave functions, we obtained variational wave functions for the three lowest exciton energy levels, $1s$, $2p$, $2s$, and also verified their accuracy for exciton calculation by checking against the RE results. Further we used these analytical wave-functions to study the Stark effects for a monolayer TMD in an in-plane electric field. We found that the ground state energy is redshifted while the $2p$ level is split into two.  We derived an analytical expression for the ground state energy shift, quadratically dependent on the field due to the second-order Stark effect, which can be conveniently used to calculate the redshift to a good accuracy.  
The numerical solution of the RE combined with the variational method provides a simple and effective approach for the study of 2D excitons in monolayer TMDs.

\begin{acknowledgments}
We acknowledge support from the Natural Science Research Funds (No. 419080500175) of Jilin University. 
\end{acknowledgments}

\newpage


\newpage

\begin{table*}[htbp]
	\begin{center}
		\leavevmode
		\setlength{\tabcolsep}{8pt}
		\renewcommand\arraystretch{1.2}
		\caption{\label{table:1} Comparison of exciton binding energies of the ground state $1s$ and the first and second excited states $2p$ and $2s$, calculated with the variational method (VM) [Eqs.~(\ref{E10})-(\ref{E20})] and from the solution of the radial equation (\ref{eigR}), for freestanding monolayer MoS$_2$ and monolayer MoS$_2$ on three different substrates SiO$_2$, hBN or diamond  with background dielectric constant $\varepsilon=(1+\varepsilon_s)/2$. }
		\begin{tabular}{c c c c c c c c c c c}
			\hline
			\hline 
			 Substrate & $\varepsilon$ & \multicolumn{3}{c}{$\vert E_{10}\vert$} & \multicolumn{3}{c}{$\vert E_{2\pm 1}\vert$}    &  \multicolumn{3}{c}{$\vert E_{20}\vert$}     \\
		&	 & RE   &   VM  & SHM  &   RE  &   VM & SHM &   RE  &   VM &SHM \\  \hline 			
vacuum  & 1 & 0.555   &   0.543  & 0.428 &   0.318    &   0.312 & 0.216  &  0.258   & 0.242  &0.243 \\ 				
SiO$_2$ & 1.55 & 0.431   &   0.422 & 0.385 &   0.217   &   0.214 & 0.168 &  0.172   & 0.161 &0.185 \\ 		
hBN & 3 & 0.269   &   0.263  &0.292 &   0.105    &   0.103  & 0.091 &  0.080   & 0.076 & 0.097 \\ 		
diamond & 3.35 & 0.246   &   0.240  & 0.273 &   0.089    &   0.089  & 0.080 &  0.070   & 0.066  &0.084\\ 		
			\hline
			\hline 
		\end{tabular}			
	\end{center}		
	
\end{table*}

\newpage

\begin{table*}[htbp]
	\begin{center}
		\leavevmode
		\setlength{\tabcolsep}{4pt}
		\renewcommand\arraystretch{0.8}
		\caption{\label{table:2}  Ground state exciton binding energies, obtained from  the solution of the radial equation (RE) (\ref{eigR}) using reduced mass $\mu$ (in $m_0$) and polarizability $\alpha_{2D}$ (in $\AA$) of Ref.\cite{Olsen:2016} and also from other calculations in previous studies (see text),  and their  experimental values for freestanding monolayer TMDs and monolayer TMDs on a SiO$_2$ or hBN substrate [dielectric constants $\varepsilon_s$(SiO$_2$)=2.1,  $\varepsilon_s$(hBN)=5 at optical frequencies].  }		
		\begin{tabular}{p{1.0cm}<{\centering} p{1.0cm}<{\centering} p{1.0cm}<{\centering} p{1.8cm}<{\centering} p{1.2cm}<{\centering} p{1.8cm}<{\centering} p{3.2cm}<{\centering} p{1.2cm}<{\centering} p{2.0cm}<{\centering} }
			\hline
			\hline 
		 TMD & $\mu$  & $\alpha_{2D}$  & Substrate  &  \multicolumn{4}{c}{$\lvert E_{10}\rvert$ (eV)}      \\
	&	 &    &    &  RE & \multicolumn{2}{c}{Other Calculation}   &  \multicolumn{2}{c}{Experiment}   \\  \cline{5-9} 			
MoS$_2$  & 0.28 & 7.1   &   vacuum  &   0.542  & 0.5\cite{Feng:2012}, & 0.54\cite{Berkelbach:2013,Olsen:2016,Shih:2013,Kylanpaa:2015},  &    & $\ge$0.57\cite{Klots:2014} \\ 	
 &   &   &    &    & 0.72\cite{Defo:2016},   & 1.0\cite{Qiudy:2013,Ramasubramaniam:2012} &  & \\ 	
  &   &   &   SiO$_2$  &   0.424  & 0.349\cite{Kylanpaa:2015}, & 0.46\cite{Berghauser:2014},  &    & 0.44\cite{Hill:2015}, \\ 	
 &   &   &    &    &  0.48\cite{Defo:2016}   & &  & 0.31\cite{Rigosi1:2016} \\ 	
  &   &   &   hBN  &   0.269  & 0.45\cite{Defo:2016} &    &  0.22$^a$\cite{Robertc:2018},  & 0.22$^b$\cite{Zhangc:2015} \\ 	
MoSe$_2$  & 0.27 & 8.15  &   vacuum  &   0.484  & 0.47\cite{Berkelbach:2013}, & 0.48\cite{Olsen:2016,Kylanpaa:2015,Mayers:2015},  &    &  \\ 	
 &   &   &    &    &  0.65\cite{Ugeda:2014},  & 0.9\cite{Ramasubramaniam:2012,Konabe:2014} &  & \\ 	
  &   &   &   SiO$_2$   &   0.382  & 0.323\cite{Kylanpaa:2015} &   &    & 0.59\cite{Liuhj:2015} \\ 	
  &   &   &   hBN   &   0.245 &  &    &    &  \\ 	
WS$_2$  & 0.22 & 6.35   &   vacuum  &   0.552  & 0.5\cite{Berkelbach:2013,Kylanpaa:2015}, & 0.54\cite{Olsen:2016,Shih:2013,Mayers:2015},  &    &  \\ 	
 &   &   &    &    &  0.66\cite{Defo:2016} ,   & 0.59\cite{Andersen:2015},1.0\cite{Ramasubramaniam:2012,Konabe:2014}&  &   \\ 	
  &   &   &   SiO$_2$   &   0.425  & 0.41\cite{Defo:2016}, &  0.323\cite{Kylanpaa:2015}  &  0.36\cite{Rigosi1:2016},   & 0.32\cite{Hill:2015,Chernikova1:2014}, \\ 	
 &   &   &    &    &   & & 0.41\cite{Stier:2016}, &0.7\cite{Yez:2014,Zhubr1:2015}  \\ 	
  &   &   &   hBN   &   0.265  & 0.37\cite{Defo:2016}, & 0.4\cite{Andersen:2015}   &    &  \\ 	
WSe$_2$  & 0.23 & 7.36   &   vacuum   &   0.501  & 0.494\cite{Olsen:2016},  & 0.46\cite{Kylanpaa:2015,Mayers:2015,Berkelbach:2013},  &    &  \\ 	
 &   &   &    &    &  0.6\cite{Defo:2016},   & 0.9\cite{Ramasubramaniam:2012,Konabe:2014}&  &   \\ 	
  &   &   &   SiO$_2$   &   0.390  & 0.295\cite{Kylanpaa:2015} &    & 0.2\cite{Huangja1:2016},   & 0.37\cite{He:2014}, \\ 	
   &   &   &      &     &  &    & 0.6\cite{Wangg2:2015},   & 0.7\cite{Liuhj:2015,Diware:2018},  \\ 
   &   &   &      &     &  &    &    & 0.89\cite{Hanbicki:2015} \\    
  &   &   &   hBN   &   0.244  &  &     &   & 0.245$^c$\cite{Poellmann:2015} \\ 	
			\hline
			\hline 
		\end{tabular}			
	\end{center}		
 $^a$Monolayer MoS$_2$ encapsulated in hBN layers. 
\\$^b$Monolayer MoS$_2$ on graphite with $\varepsilon_s$=7.
\\$^c$Monolayer WSe$_2$ on diamond with $\varepsilon_s$=5.7.
	
\end{table*}

\newpage

\begin{figure}

\caption
{(Color online) Exciton spectrum of freestanding monolayer MoS$_2$ calculated from the solution of the radial equation (RE) (\ref{eigR}). Effective Bohr radii of six low lying  exciton energy levels are given in the brackets (in $\AA$). 
}
\label{fig1}
\vspace*{5mm}

\caption
{(Color online) Radial functions $R_{nm}(r)$ of the six low-energy exciton states, (a) $1s$, $2s$, $3s$ and (b) $2p$, $3p$, $3d$, calculated from the radial equation (RE) (\ref{eigR}).   
}
\label{fig2}
\vspace*{5mm}

\caption
{(Color online) Exciton Rydberg series $\lvert E_{nm}\rvert$ versus principle quantum number $n$, for the three smallest values of orbital quantum number (a) $\lvert m\rvert=0$, (b) $\lvert m\rvert=1$, (c) $\lvert m\rvert=2$, of an isolated monolayer of MoS$_2$ calculated with the screened hydrogen model (SHM) [Eqs.~(\ref{epseff12}) and (\ref{exoneff})] and from the solution of the radial equation (RE) (\ref{eigR}). The effective dielectric constants $\varepsilon_{nm}$ versus $n$ of the SHM [Eq.~(\ref{epseff12})] for the three $\lvert m\rvert$ values are shown in the insets.    
}
\label{fig3}
\vspace*{5mm}

\caption
{(Color online) Exciton effective radii $a_{nm}$ versus principle quantum number $n$ for excitons $s$ ($\lvert m\rvert=0$), $p$ ($\lvert m\rvert=1$), $d$ ($\lvert m\rvert=2$) in freestanding monolayer MoS$_2$, calculated from (a) the radial equation (RE) (\ref{eigR}) and (b) the screened hydrogen model (SHM) [Eqs.~(\ref{epseff12}) and (\ref{exoneff})]. 
}
\label{fig4}
\vspace*{5mm}

\caption
{
(Color online) Exciton binding energies $\lvert E_{nm}\rvert$ versus orbital quantum number $m$ of freestanding  monolayer MoS$_2$, for principal quantum number $n$=2-6 calculated (a) from the radial equation (\ref{eigR}) and (b) with the screened hydrogen model (SHM) [Eqs.~(\ref{epseff12}) and (\ref{exoneff})].  
}
\label{fig5}
\vspace*{5mm}

\caption
{
(Color online) Exciton effective radii $a_{nm}$ versus orbital quantum number $m$ of  freestanding monolayer MoS$_2$, for principal quantum number $n$=2-6 calculated from (a) the radial equation (RE) (\ref{eigR}), (b) the screened hydrogen model (SHM) [Eqs.~(\ref{anmeff}) and (\ref{epseff12})] and (c) the 2D hydrogen model (2DHM) [Eq.~(\ref{anmeff}) with $\varepsilon_{nm}=1$].  
}
\label{fig6}

\end{figure}

\newpage

\begin{figure}

\caption
{(Color online) Expectation value of (a) $E_{10}$ of the ground state ($1s$) exciton versus variational parameter $a$  [expression~(\ref{E10})], (b) $E_{21}$ of the first excited state ($2p$) exciton versus variational parameter $a'$  [expression~(\ref{E21})] and (c) $E_{20}$ of the second excited state ($2s$) exciton versus variational parameter $a''$ [expression~(\ref{E20})] after finding the variational parameter $a=10.4~\AA$ from (a) for an isolated MoS$_2$ monolayer. 
}
\label{fig7}
\vspace*{3mm}

\caption
{(Color online) Radial functions of the exciton (a) ground state $1s$, (b) first excited state $2p$ and (c) second excited state $2s$ as obtained from the variational method and the solution of the radial equation (RE) (\ref{eigR}) for a freestanding   monolayer of MoS$_2$.  
}
\label{fig8}
\vspace*{3mm}

\caption
{(Color online)  Exciton Rydberg series $E_{nm}$ versus principle quantum number $n$ of monolayer WS$_2$ on the SiO$_2$ substrate ($\varepsilon_s=2.1$)  for (a) the $s$ ($m$=0), $p$ ($m$=1), $d$ ($m$=2) exciton states,  calculated from the radial equation (RE) (\ref{eigR}) and also with the variational method (VM) [expressions~(\ref{E10}), (\ref{E21}) and (\ref{E20})]  for excitons $1s$, $2p$, $2s$ (solid squares), and (b) the $s$ ($m$=0) exciton states, calculated from the radial equation (RE)  (\ref{eigR}), the screened hydrogen model (SHM) and 2D hydrogen model (2DHM) [expression (\ref{E2dcoul})] and obtained from the reflectance contrast measurements in the  experimental study \cite{Chernikova1:2014}.  Reduced mass $\mu=0.22m_0$ and 2D polarizability $\alpha_{2D}$=6.35 $\AA$ from DFT calculations of Ref.\cite{Olsen:2016}  were used in our calculation. The 2DHM yields a $1s$ exciton energy $\sim$-5 eV that   is too low to be shown in (b). 
}
\label{fig9}
\vspace*{5mm}

\caption
{(Color online)  Exciton energies $E_{nm}$ versus orbital quantum number $m$ of monolayer WS$_2$ on a SiO$_2$ substrate, for principal quantum number $n$=2-5 calculated from the radial equation (\ref{eigR}). The material parameters  are the same as in Fig.~\ref{fig9}. 
}
\label{fig10}
\vspace*{5mm}

\caption
{(Color online)  Exciton energies versus strength of an in-plane electric field for (a) the ground state $1s$ and (b) the excited states $2p$, $2s$ of freestanding monolayer  MoS$_2$, which are the solutions to Eq.~(\ref{seqfild}), obtained with a model of four exciton states that are described at zero field by the four variational wave functions  [Eqs.~(\ref{fi00})-(\ref{fi10})].  The dotted curve in (a) represents the quadratic dependence of the ground state exciton energy on the field strength (see text for the analytical expression), obtained by a second-order perturbation theory.  Strictly speaking the notation of $1s$, $2p$, $2s$ is used for the exciton states at zero field. 
}
\label{fig11}

\end{figure} 


\newpage

\begin{figure*}
	\centering
	\includegraphics[width=1.2\textwidth]{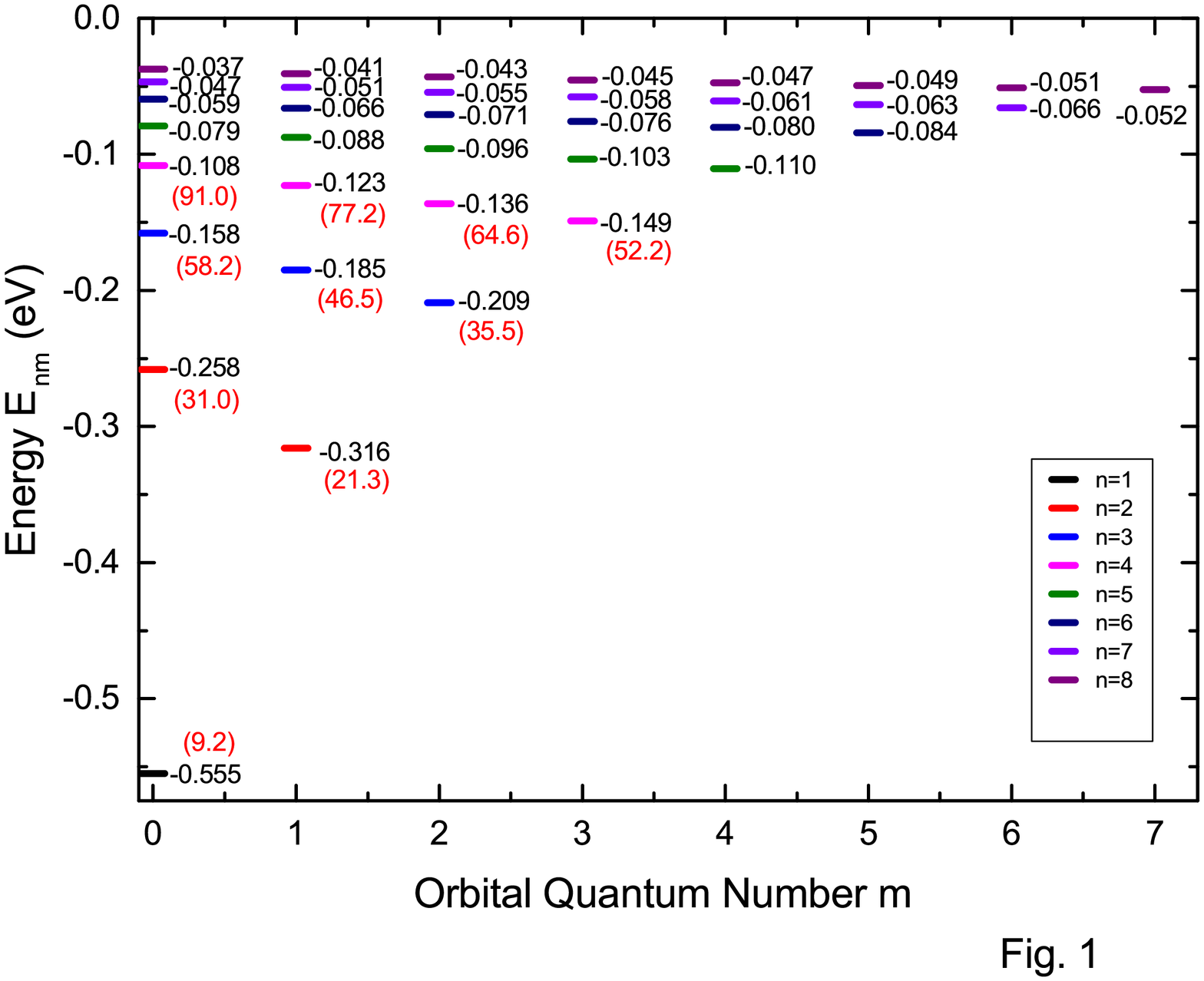}
\end{figure*}

\newpage

\begin{figure*}
	\centering
	\includegraphics[width=0.9\textwidth]{fig2vv.eps}
\end{figure*}

\newpage

\begin{figure*}
	\centering
	\includegraphics[width=0.8\textwidth]{fig3vv.eps}
\end{figure*}

\newpage

\begin{figure*}
	\centering
	\includegraphics[width=0.9\textwidth]{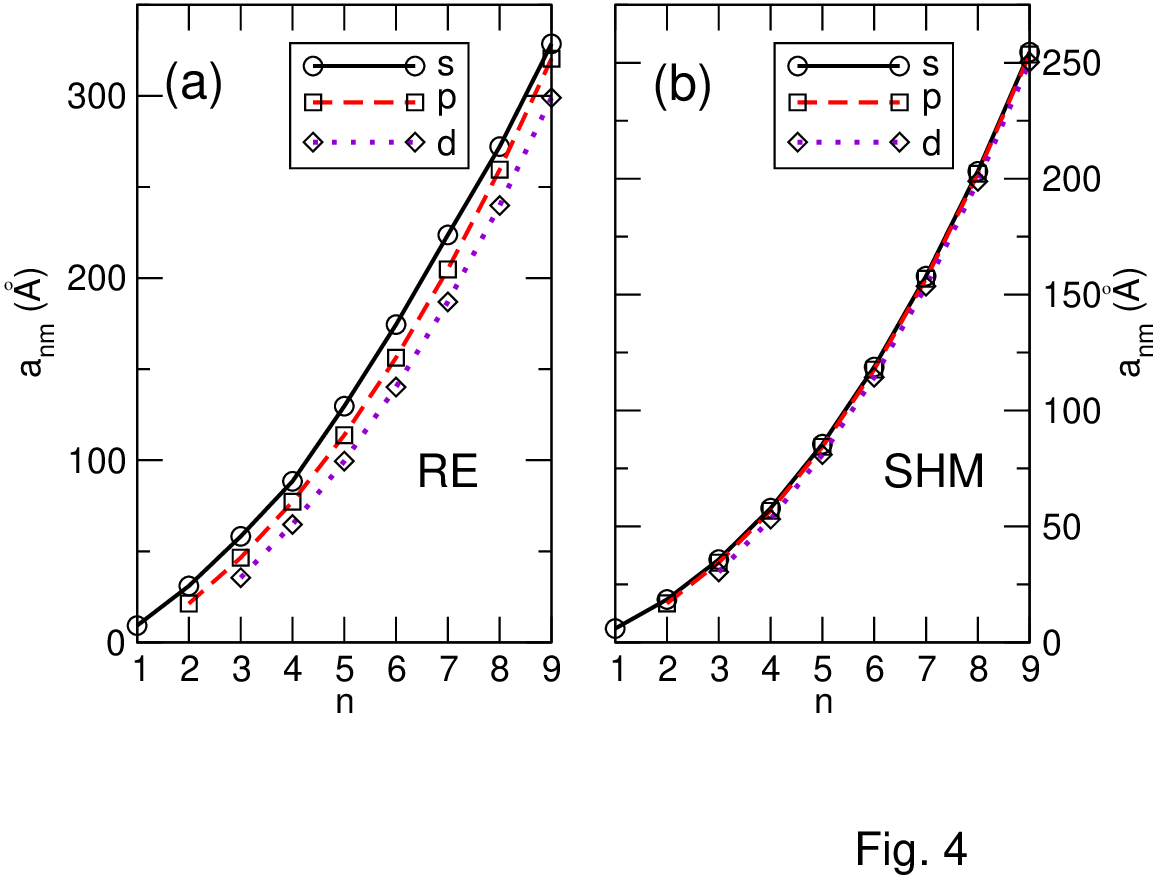}
\end{figure*}

\newpage

\begin{figure*}
	\centering
	\includegraphics[width=1.0\textwidth]{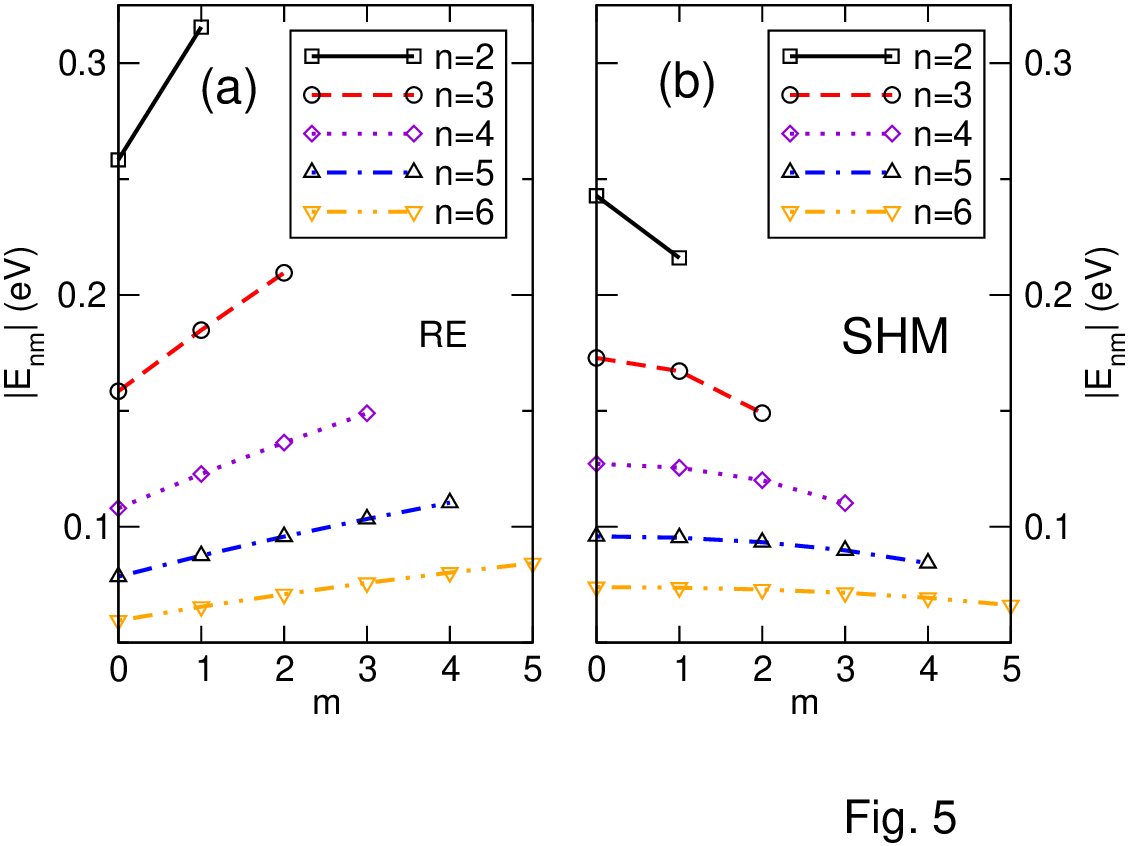}
\end{figure*}

\newpage

\begin{figure*}
	\centering
	\includegraphics[width=1.0\textwidth]{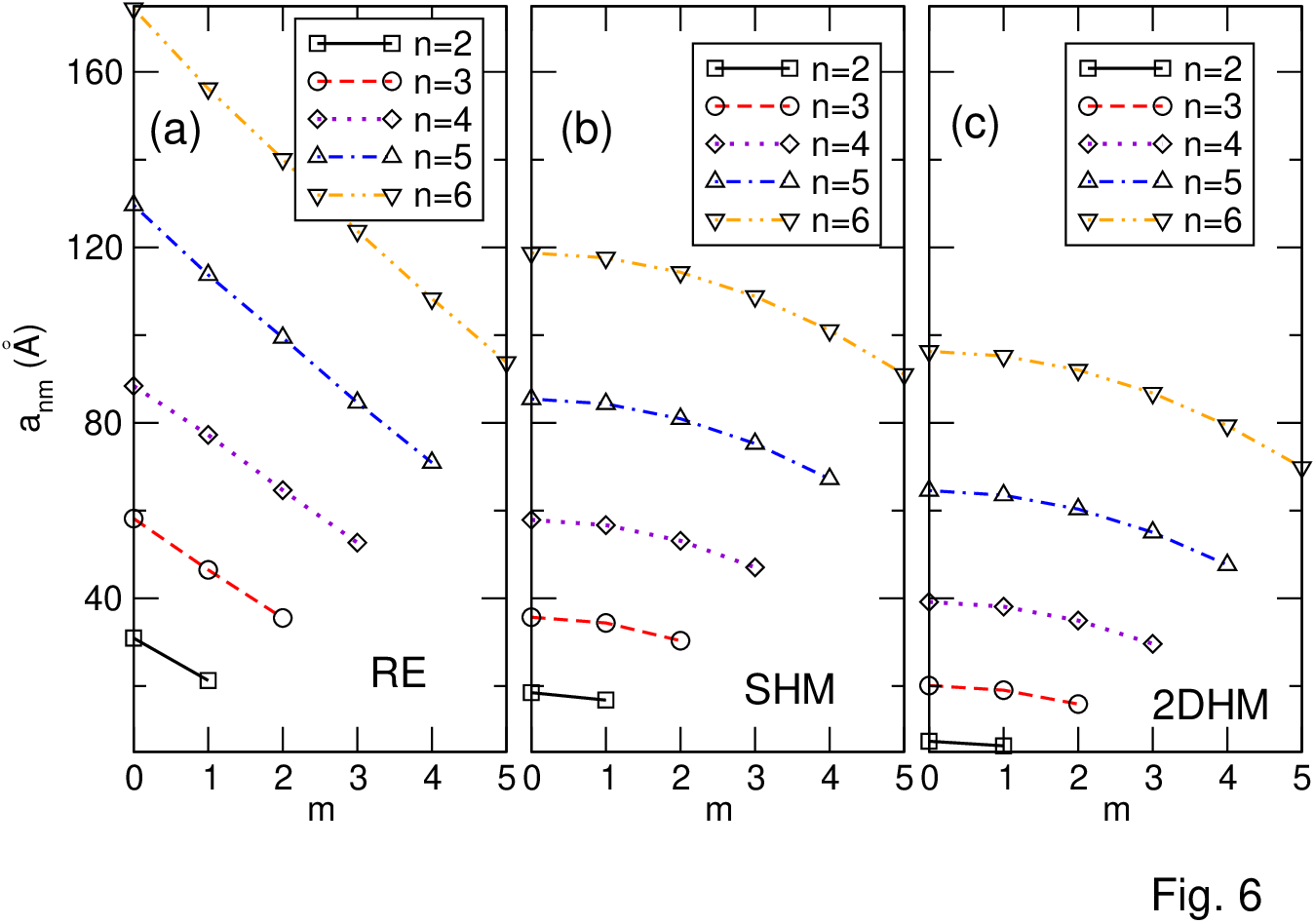}
\end{figure*}

\newpage

\begin{figure*}
	\centering
	\includegraphics[width=0.8\textwidth]{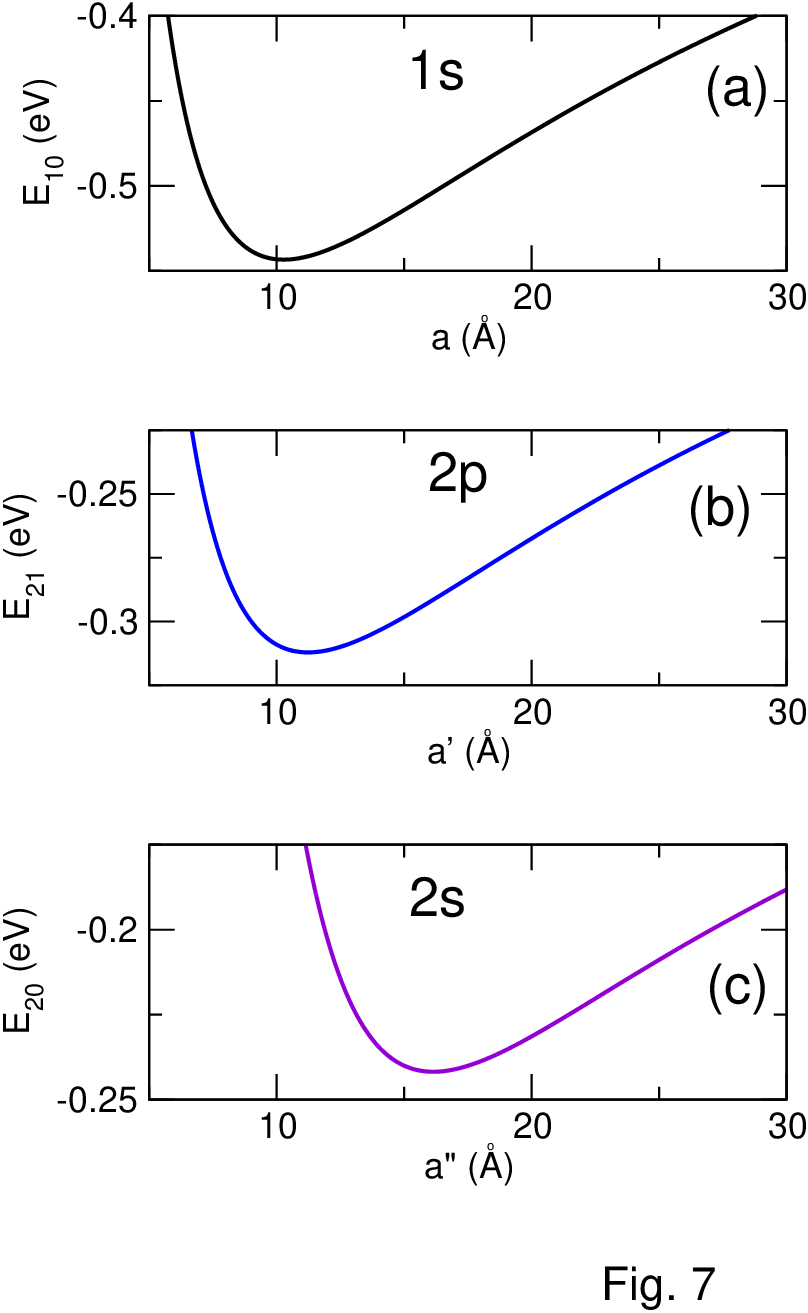}
\end{figure*}

\newpage

\begin{figure*}
	\centering
	\includegraphics[width=0.8\textwidth]{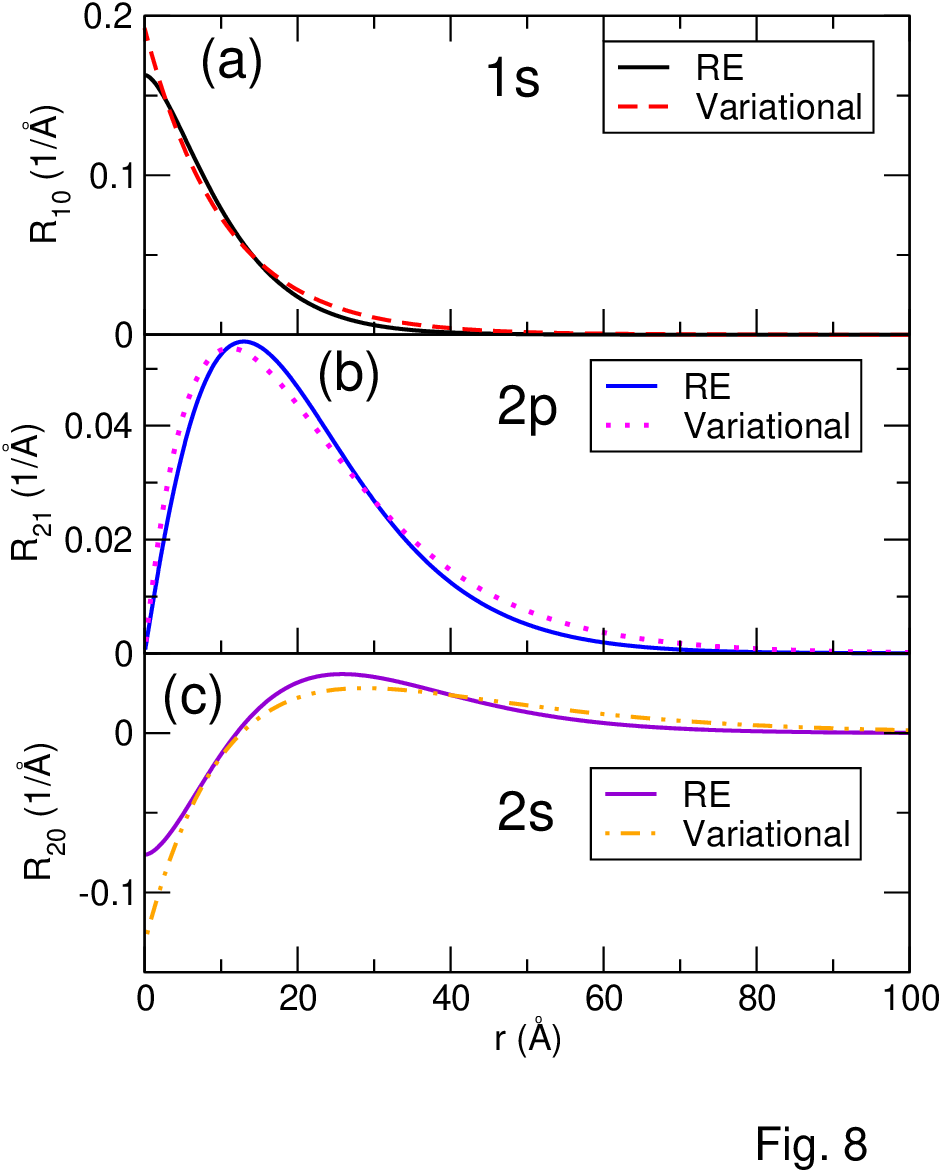}
\end{figure*}

\newpage

\begin{figure*}
	\centering
	\includegraphics[width=0.9\textwidth]{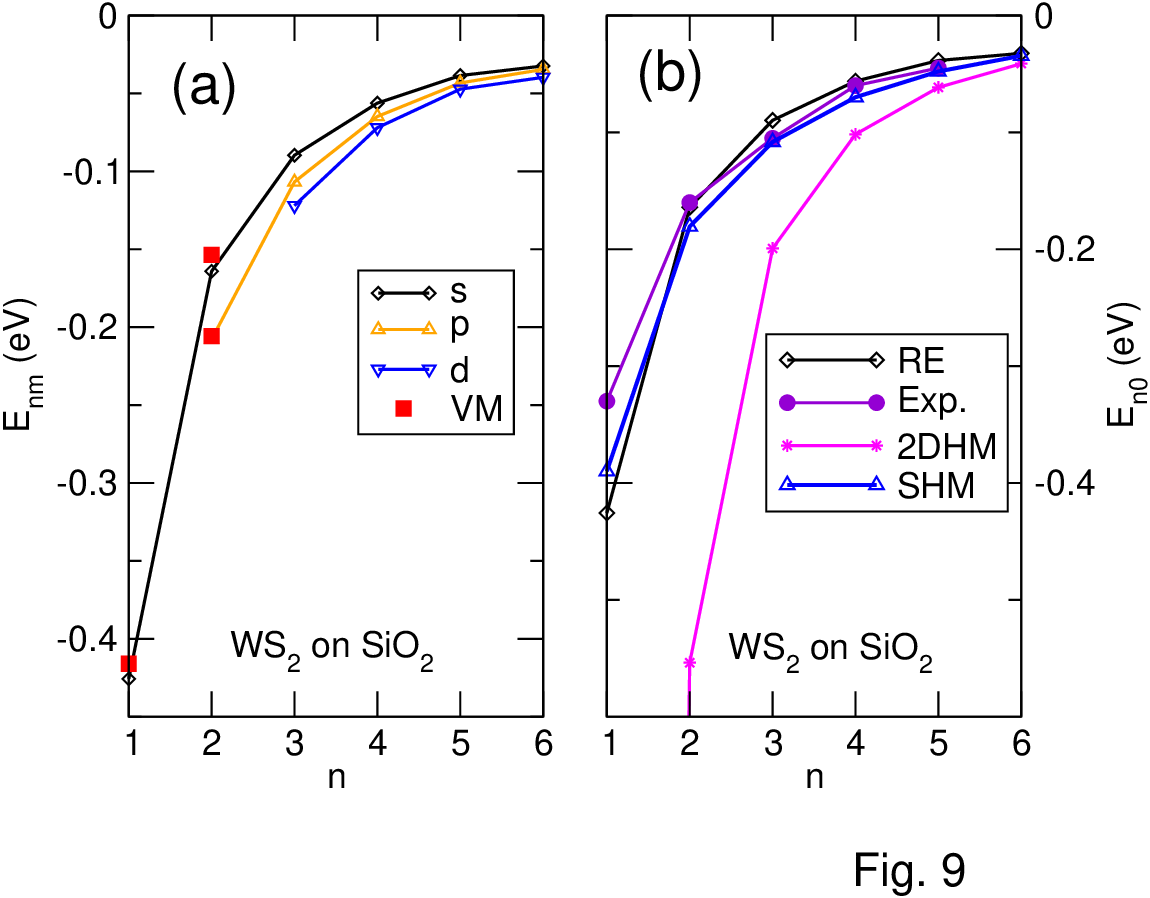}
\end{figure*}

\newpage

\begin{figure*}
	\centering
	\includegraphics[width=0.8\textwidth]{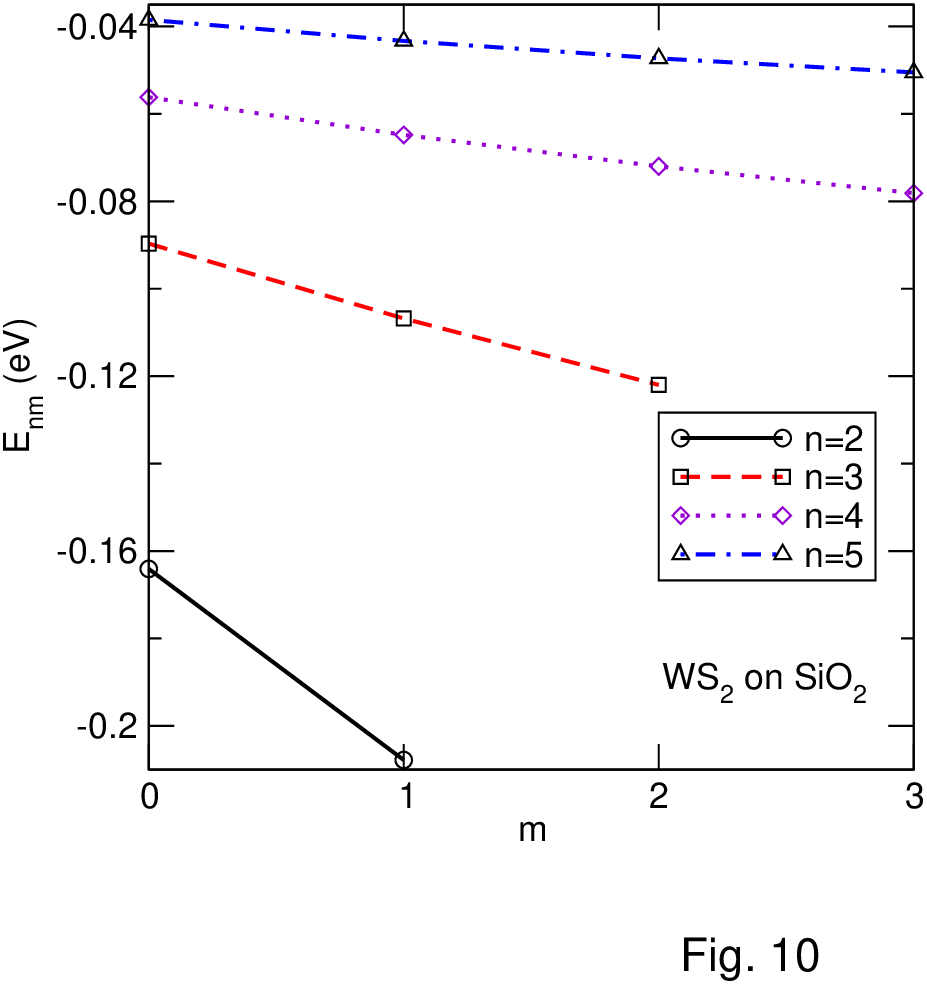}
\end{figure*}

\newpage

\begin{figure*}
	\centering
	\includegraphics[width=0.9\textwidth]{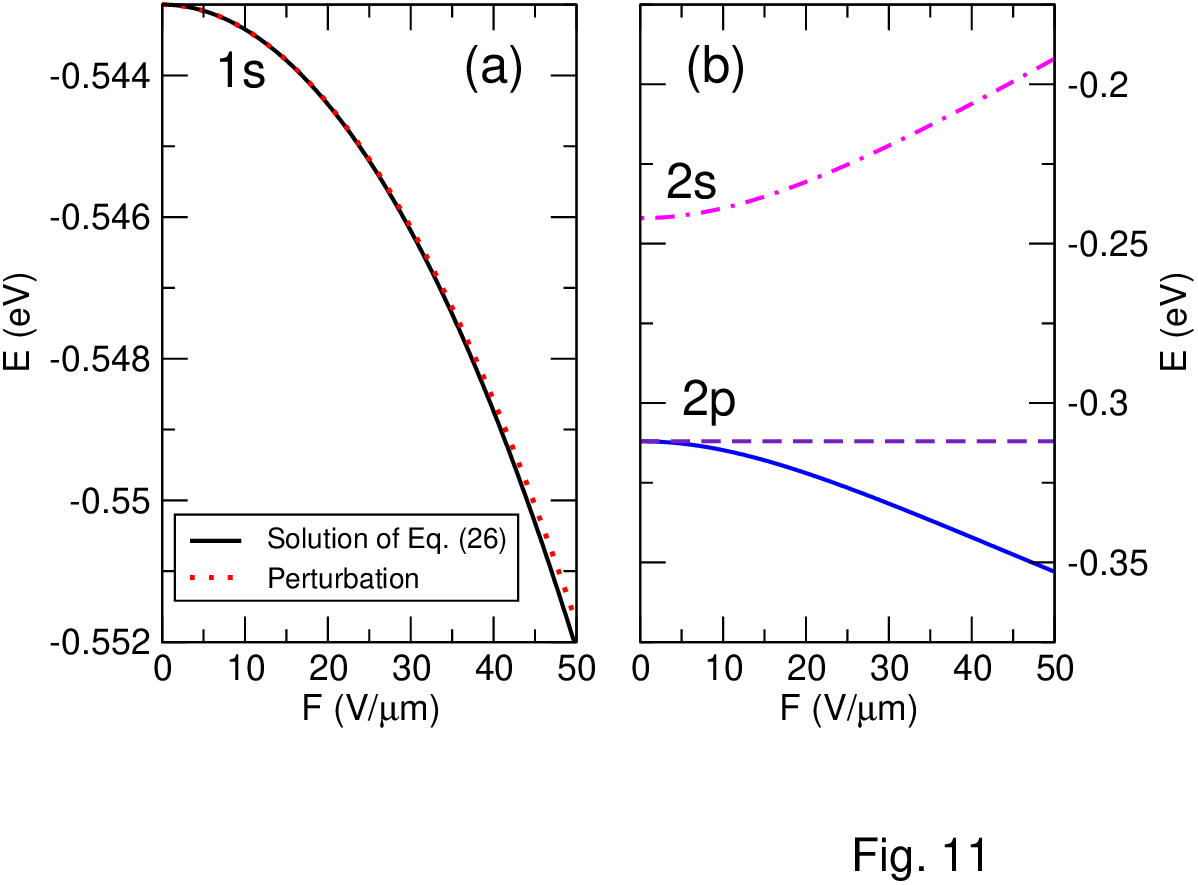}
\end{figure*}

\end{document}